\documentclass[11pt]{article}
\usepackage[OT2, T1]{fontenc}
\usepackage[english]{babel}
\usepackage{authblk}
\usepackage{epsf}
\usepackage[dvips]{graphics,graphicx}
\usepackage{cite}
\usepackage{enumerate}
\usepackage[normalem]{ulem}
\usepackage{amsmath,amssymb,bm,color,mathrsfs,wasysym}
\usepackage{epsfig}
\usepackage{amsfonts}
\usepackage{cancel}
\usepackage{float}
\usepackage{perpage}
\usepackage{setspace}
\usepackage{nameref}
\usepackage{hyperref}
\hypersetup{colorlinks}
\usepackage{xcolor}
\usepackage[lmargin=3cm,rmargin=3cm,tmargin=2cm,bmargin=2cm,marginpar=2cm ,reversemp]{geometry}

\begin{document}
	
	\title{\textbf{Finite-temperature avalanches in 2D disordered Ising models}}

	\author[1]{Federico Ettori
	\thanks{federico.ettori@polimi.it}}
	\author[2]{Filippo Perani
	\thanks{filippo.perani@polimi.it}}
	\author[2]{Stefano Turzi
	\thanks{stefano.turzi@polimi.it}}
	\author[1]{Paolo Biscari
	\thanks{paolo.biscari@polimi.it}}

	\affil[1]{\small Department of Physics, Politecnico di Milano, Piazza Leonardo da Vinci 32, 20133 Milan, Italy}
	\affil[2]{\small Department of Mathematics, Politecnico di Milano, Piazza Leonardo da Vinci 32, 20133 Milan, Italy}
	\date{}

	\maketitle
	
	\begin{abstract}
		We study the qualitative and quantitative properties of the Barkhausen noise emerging at finite temperatures in random Ising models. The random-bond Ising Model is studied with a Wolff cluster Monte-Carlo algorithm to monitor the avalanches generated by an external driving magnetic field. Satisfactory power-law distributions are found which expand over five decades, with a temperature-dependent critical exponent which matches the existing experimental measurements. We also focus on a Ising system in which a finite fraction of defects is quenched. Also the presence of defects proves able to induce a critical response to a slowly oscillating magnetic field, though in this case the critical exponent associated with the distributions obtained with different defect fractions and temperatures seems to belong to the same universality class, with a critical exponent close to 1.
	\end{abstract}
	
	\section{Introduction} \label{sec:introduction}
	
	The Barkhausen noise (BN) is a classical physical phenomenon that manifests itself as a series of magnetization jumps in ferromagnetic samples as  they reverse their spins under the action of a varying external magnetic field \cite{19Barkhausen,96Stanley,00Puppin,21metra}. It bears both practical and theoretical significance, as it emerges as a macroscopic manifestation of the bursty evolution at the microscopic domain scale. One of the most prominent features of BN is that the amplitude $\Delta M$ of the magnetization jumps follows a power-law probability distribution $P(\Delta M)\sim (\Delta M)^{-\tau}$, with $\tau$ critical exponent \cite{95Sethna,00Puppin,03Zani,04Puppin}.
	
	Different theoretical approaches have been proposed to explain BN \cite{zapbook}. It has been ascribed to the presence of defects or local impurities which act as pinning sites for the magnetic domain walls. This is a dynamic effect where the domain walls evolve in random walks, and generate discrete magnetization jumps which drive the system across metastable states \cite{98zapdur}. Another common theoretical explanation is related to the concept of self-organised criticality \cite{87Bak,88Bak,01Sethna}. In broad terms, criticality appears when the microscopic interaction between parts of the system is sufficiently strong, but not very strong, and, at the same time, the internal disorder is sufficient to generate regions of weak interaction \cite{01Sethna}. The yielding events can thus span many size scales. Within a many-degree-of-freedom model approach, it has been argued \cite{93Sethna,95Sethna} that the zero-temperature random-field Ising model provides a universal, quantitative explanation for experimental results.
	
	Most of the models for BN presented until now refer to zero or very low \cite{21metra} temperature, whereas BN can be observed at room temperature also. The role of temperature is discussed only in few experimental or theoretical works, also within the framework of other physical phenomena exhibiting self-organised criticality \cite{87Bak,95Vergeles,98Caldarelli,04Puppin}.
	
	A simple way of considering the competing effects of ordering interactions and defects, at various temperatures, is to introduce some kind of random disorder on simple models. Among the various possibilities, we here consider two disorder-enriched variants of the classical Ising model. The first is the well-known random-bond Ising model (RBIM) \cite{94Vives,95Vives,Fytas2010}, where the interaction strength between neighbouring spins is a random variable with positive mean. We also consider an Ising model in the presence of defects, modelled as fixed spins with random orientation \cite{22esb}. Both models exhibit a disorder-induced phase transition and magnetic avalanches with power-law distributions which mimic the statistics of Barkhausen noise.
	
	At variance with several previous studies, we consider the finite temperature case and study the temperature dependence of the critical exponent. The measured power-law critical exponent $\tau$ is in line with experimental observations, but the two models exhibit different temperature behaviour. In the RBIM the temperature effect is significant, and $\tau$ changes (approximately) linearly from $\approx 1$ at temperatures close to the Curie point, to $\approx 1.6$ at low temperature, quite in agreement with the experimental results of Puppin and Zani \cite{00Puppin,04Puppin}. On the contrary, the pinned-defects model exhibits a much weaker temperature dependence of the power-law exponent, which remains $\tau \approx 1$ in the whole range of explored temperatures, thus suggesting that both Ising variants do not belong to the same universality class.
	
	The paper is organised as follows. In Sect.~\ref{sec:MC} we remind the basic properties of the RBIM, and present the adapted version of the Wolff cluster Monte-Carlo algorithm we use to monitor the Barkhausen jumps. In Sect.~\ref{sec:Para-ferro} we use the above algorithm to confirm the presence of a complex phase diagram, including para, ferro, and glassy phases, and to derive the critical exponent characterising BN. In Sect.~\ref{sec:def} we study the behaviour of the Ising model with random defects. The crucial role of the measurement time is stressed, and the relevant critical exponent is derived at different temperatures. In the concluding section we analyse finite size effects and discuss our results.
	
	\section{Cluster simulation of the Random-Bond Ising Model}
	\label{sec:MC}
	
	We consider a two-dimensional $L \times L$ square lattice, where each site is occupied by a spin variable $s_i$ ($i =1, \ldots , N=L^2$), taking values $\pm 1$. The Hamiltonian is
	\begin{equation}
		H[\textsf{s}] = -\sum_{\langle i,j \rangle} J_{ij} s_i s_j - B \sum_i s_i,
		\label{eq:Hamiltonian}
	\end{equation}
	where $\langle i,j \rangle$ denotes a nearest-neighbour pair of spins, $J_{ij}$ is the interaction strength and $B$ is an external magnetic field. In the RBIM the interaction strengths are chosen at random according to a normal distribution with (positive) mean $\bar J$, and variance $R$. In the following, we set $\bar J=1$ so that the temperatures will be measured in units of $\bar J/k_B$, with $k_B$ the Boltzmann constant. Notice that for any non-vanishing value of the \emph{disorder parameter} $R$ there is a non-vanishing probability of having $J_{ij}<0$, \emph{i.e.}, antiferromagnetic interactions may be present in the sample among random pairs of spins.
	
	Most studies in this context make some further simplifying assumptions. For example, the analysis is performed only at $T=0$, or fully connected spin networks (mean-field model) are considered \cite{82Hemmen,95Vives,07Nogueira,Janke}. When short range interactions are taken into account, a typical model is the $\pm J$ model, which has antiferromagnetic bonds $-J$ with probability $p$, and ferromagnetic bonds $+J$ with probability $1-p$. On a square lattice, when $p$ exceeds a critical value $p_c$, it is expected to observe a transition from a ferromagnetic to a spin-glass phase \cite{94Vives,07Nogueira}.
	
	We want to study the evolution of the system \eqref{eq:Hamiltonian} when the external field $B$ is slowly changed. For every step of the external field we record the dynamics of the magnetic domains' reversal. As a result we will find that at sufficiently high disorder $R$, magnetization avalanches emerge, with an intensity obeying a power-law distribution which spans over several orders of magnitude.
	
	In order to monitor the intensity of the magnetization avalanches, the Wolff cluster algorithm emerges as the natural choice for the simulations, as it allows the domain formation, and promotes the reversal of domains possibly of any size, which is at the basis of the physical intuition behind the Barkhausen effect. Indeed, the cluster algorithm by construction generates magnetic domains and avoids the critical slowing down typically observed in single-spin-flip algorithms. The cluster algorithm must be however suitably adapted to account for the presence of a variable external field and the possible presence of antiferromagnetic bonds. First of all, and since the cluster algorithm is non-local \cite{98jan,18seth}, extra care must be taken in the presence of variable external fields, as we are allowed to trust the predictions of the Cluster algorithm only in configurations of (quasi-)equilibrium. This impacts the choice of the speed at which the driving field can be varied, and in particular implies that the driving field variation must be sufficiently slow. The magnetic field is taken into account by introducing a ghost-spin $s_0$ \cite{67Griffiths,89Wang}, which is assumed to be connected to all the other $N$ spins in the lattice. It takes the sign of the external field so that the Hamiltonian can be rewritten as
	\begin{equation}
		H[\textsf{s}] = -\sum_{\langle i,j \rangle} J_{ij} s_i s_j - \lvert B \rvert \sum_{i=1}^{N} s_0 s_i = -\sum_{\langle i,j \rangle} \hat{J}_{ij} s_i s_j,
		\label{eq:Hamiltonian_ghost}
	\end{equation}
	where we remark that in the final expression the indices $i,j$ range between $0$ (the ghost spin) and $N$, and
	\begin{equation}
		\hat{J}_{ij} =
		\begin{cases}
			J_{ij} & \text{ if $i$ and $j$ are neighbours and } i, j > 0, \\
			\lvert B \rvert & \text{ if } i=0 \text{ or } j = 0.
		\end{cases}
	\end{equation}
	The possible presence of antiferromagnetic bonds will be dealt within the point~\eqref{antiferro} described below.
	
	We monitor the state of the system through the average magnetization $M=1/N \sum_i s_i$, and the magnetic susceptibility, which we derive from the fluctuation-dissipation theorem
	\begin{equation}
		\chi=\frac{\langle M^2\rangle-\langle M\rangle^2}{k_B T},
	\end{equation}
	where $\langle \,\cdot \,\rangle$ denotes the ensemble average.
	
	Clusters are then constructed by using this new lattice with coupling strengths $\hat{J}_{ij}$. Furthermore, the presence of randomness and antiferromagnetic bonds is taken into account by suitably modifying the adding probability for the cluster growth. Specifically, the algorithm proceeds as follows.
	\begin{enumerate}
		\item \label{step1} Choose a seed spin, say $s_i$, at random. Put it first on a list which enumerates the sites belonging to the growing cluster.
		\item \label{step2} Pick $s_k$, the first unvisited spin from the cluster list (at the first iteration this will be $s_i$).
		\begin{enumerate}
			\item \label{antiferro} Look in turn at each of the neighbours of $s_k$. Let $s_j$ be the selected neighbour. If not already in the cluster, add $s_j$ at the end of the cluster list, with probability ($\beta = 1/k_B T$)
			\begin{equation}
				p_{\text{add}} = \max\left\{0, 1-e^{-2\beta \hat{J}_{kj}s_k s_j} \right\}.
			\end{equation}
			Note that we do not check whether $s_k$ and $s_j$ have the same orientation. By contrast with the standard Wolff algorithm, we consider also spins with opposite sign, provided that $\hat{J}_{kj}s_k s_j > 0$. In so doing, we properly account for possible antiferromagnetic interactions among the spins.
			
			\item If the ghost spin happens to be added to the cluster, stop the construction of the cluster because the ghost spin can never flip. In this case the cluster is pinned and we return to step \eqref{step1} to build a new cluster.
			
			\item Visit all the neighbours of $s_k$ and then repeat step \eqref{step2} with the next spin in the cluster list until the end of the cluster list is reached, i.e., there are no more spins left in the cluster whose neighbours have not been considered for inclusion in the cluster.
		\end{enumerate}
		
		\item Flip the cluster.
	\end{enumerate}
	
	It is easy to see that the resulting Markov chain is irreducible (there is a non-vanishing probability to reach any configuration from every other in a finite number of steps), and aperiodic. Furthermore it satisfies detailed balance. Hence, it is ergodic and converges to the Maxwell-Boltzmann canonical distribution.
	
	\section{Barkhausen noise in the RBIM}
	\label{sec:Para-ferro}
	
	The Wolff algorithm is particularly efficient close to the transition temperature $T_c$. In the presence of disorder, the critical temperature is reduced, with $T_c(R)$ a (decreasing) function. We perform our analysis close to the critical temperature, as the Wolff algorithm loses efficiency at low temperatures when random bond interactions are considered. This is because domain walls along weak bonds effectively pin the building of the cluster. As a result, only very small or very large clusters are built at low temperatures, and this critically delays the attainment of equilibrium \cite{90Kessler}.
	
\begin{figure}
		\begin{center}
			\includegraphics[width=0.75\textwidth]{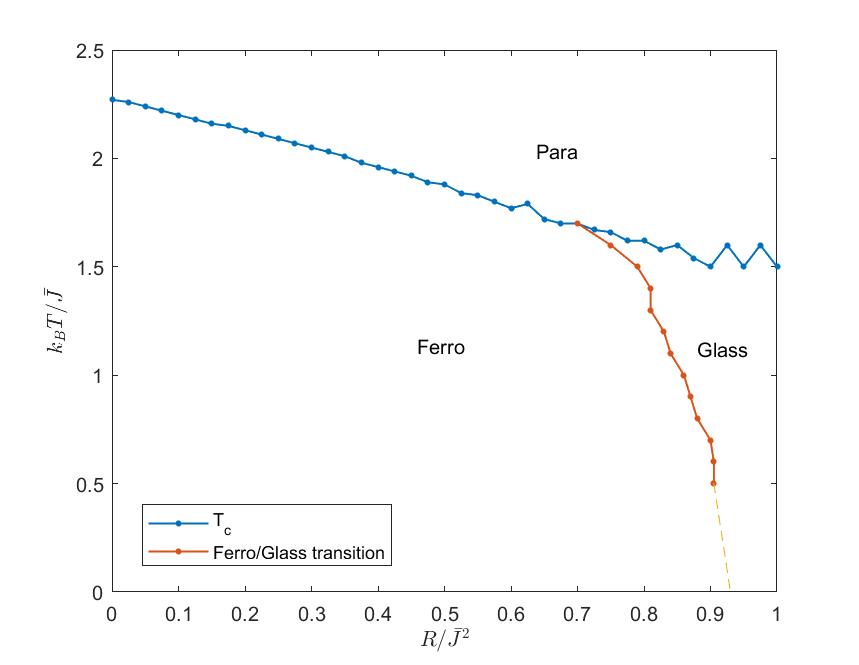}
		\end{center}
		\caption{Phase diagram of a 2D RBIM with variable degree of disorder}
		\label{fig:FerroParaGlass}
	\end{figure}

	In order to identify the temperature range best suited for our analysis, we first carry out a numerical study to understand how the critical temperature $T_c$ depends on $R$. The results of this numerical analysis are reported in Fig.~\ref{fig:FerroParaGlass}. When $R = 0$, we obtain the classical critical temperature of the Ising model without disorder, $T_c(0) \approx 2.27$. Interestingly, when $R>0$, we find an almost linear relationship between $T_c$ and $R$. The critical temperature decreases with increasing disorder $R$, which confirms the intuition that increasing the disorder is somewhat equivalent to increasing the temperature. The transition temperature is estimated by looking at the susceptibility peak, for various temperatures on a system with $N=500\times 500$ spins. Close to the transition temperature, the algorithm reaches equilibrium very quickly so that we do not need to discard many initial MC steps.
	
	The behaviour of the system varies significantly when the disorder parameter $R$ is increased above a critical value. In the disordered regime the ferromagnetic phase becomes unstable, the free energy exhibits many metastable states and a glassy-like phase emerges \cite{82Hemmen,07Nogueira}. We located the ferro-glass transition by creating $N_\text{r}=100$ different replicas of a system with specific values of $R$ and $\bar{J}$, and running a Monte-Carlo simulation at each different temperature, to compute the average magnetization. For each couple of replicas $\alpha$ and $\beta$ we then computed the overlap $q_{\alpha\beta}(T)=(1/N)\sum_i m_i^{(\alpha)}m_i^{(\beta)}$, and then monitored the overlap probability distribution $P(q;R,T)$ at each temperature. The red line in Fig.~\ref{fig:FerroParaGlass} identifies the critical disorder at which the probability distribution becomes different from zero, so that $P(0;R,T)>0$ for $R>R_\text{c}(T)$. The intersection between the red and blue lines in Fig.~\ref{fig:FerroParaGlass} locates the Nishimori point \cite{81nish,01Gruzberg}, where the para, ferro and glassy phases coexist.
	
	The need of averaging over a large number of replicas impacts significantly the computational time, especially when it turns to understand whether a system is in a ferromagnetic or in a glassy phase. For this reason the red transition line in Fig.~\ref{fig:FerroParaGlass} was obtained by studying a smaller system, composed by $90\times 90$ spins, and we henceforth restrict our analysis to disorder values at the left of this line, where only para- and ferromagnetic states are expected to emerge.
	We postpone to a future paper the detailed analysis of the glassy phase triggered by larger disorder values
	both in the present RBIM model and in the Ising model with defects. We however notice that ferro-glassy transition data in Fig.~\ref{fig:FerroParaGlass} are fully compatible with the zero-temperature available predictions, see the red dotted line which crosses the $T=0$ axis in the point predicted in \cite{jaggi80}.
	
	Once identified the most suitable temperature and disorder range for the para-ferromagnetic transition, we now proceed to study the avalanche statistics, namely, the Barkhausen noise generated during a magnetic transition between two oppositely magnetized states, driven by a steadily increasing external magnetic field.
	
	The domains in our simulations all start with all the spins pointing down, and flip up as the net force on them becomes positive. Fluctuations aside, this may occur either because $B$ increases sufficiently (thus generating a new avalanche), or because one of their neighbours flipped up and propagates an existing avalanche. When the disorder $R$ is large too many interaction strengths $J_{ij}$ are small (or even negative), so that small domains tend to be formed, and spins are prone to flip independently. On the contrary, if the disorder parameter is too small compared to $\bar{J}$, domains are likely to be large (they are certainly so at low $T$) and we will mostly observe large avalanches, or even a unique avalanche of the size of the entire system. In between, either because we change the temperature or the disorder, we might get crackling noise. At the critical disorder $R_c$ we expect to find avalanches at all scales.
	
	The lattice size for all the simulations is set to $L = 500$, and the external magnetic field is increased by steps of $\Delta B = 10^{-5}\bar{J}$, where we recall that the average interaction energy is set to $\bar{J} = 1$ so that it fixes the unit of measure for energy (and magnetic field). It is important to observe that the minimum size for the magnetization avalanches we record is basically limited by these two choices. On the one hand, avalanches of size smaller than $\Delta M=2/N$ are prevented by construction, as the minimum magnetization step is the flipping of a single spin. On the other hand, increasing the external field naturally induces a magnetization variation, and this adds up to the intermittent avalanche evolution. Using the largest possible systems and reducing the magnetic field steps sizes is then essential to broaden the magnetic avalanches' distributions obtained. With the very small step-size $\Delta B$ we choose, the thermalisation time turns out to be negligible. Moreover, albeit in the very first iterations the system is slightly out-of-equilibrium, this situation is closer to the real experimental setting, where the magnetic domains flip over under a slow but continuous tuning of the external magnetic field. As discussed in the previous section, this is an essential requirement which allows us to identify the magnetization avalanches we measure with the physical domain flipping which originates BN.
	
\begin{figure}
		\begin{center}
			\includegraphics[width=0.49\textwidth]{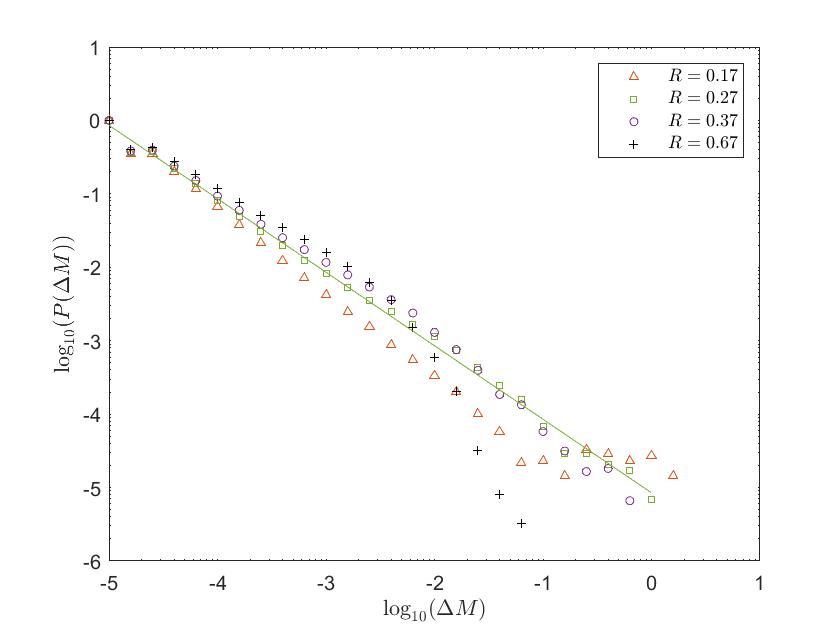}
			\includegraphics[width=0.49\textwidth]{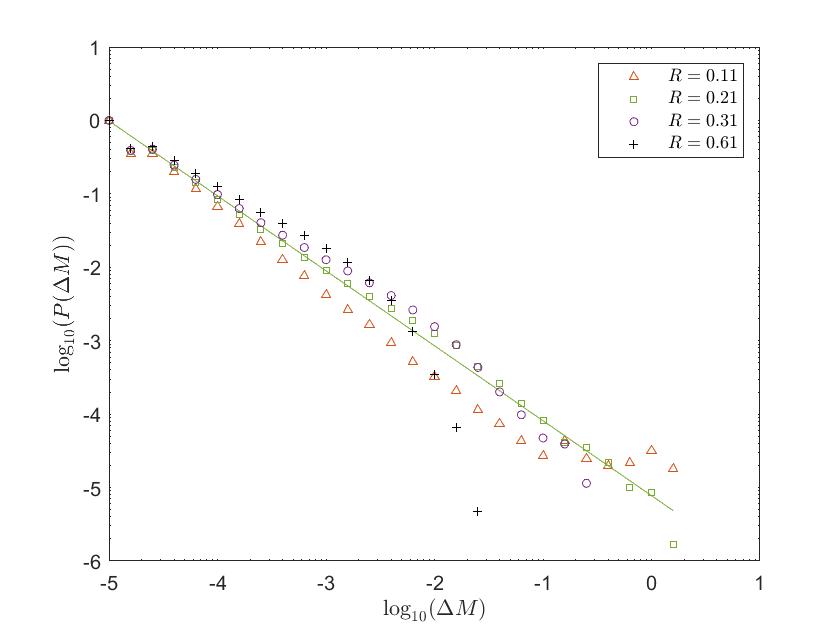}
		\end{center}
		\caption{Avalanches’ statistics for $T = 2.15$ in the left panel and $T=2.20$ in the right panel. The data refer to different values of the disorder parameter $R$: orange ($R = 0.17$) points (triangles) are sub-critical. Green ($R = 0.27$) points (squares) exhibit the longest power-law distribution, while purple ($R=0.37$)  and black ($R=0.67$) points (circles and crosses) are supra-critical.}
		\label{fig:fig03}
	\end{figure}

	The amplitude of an avalanche is measured by the variation $\Delta M$, where $M$ is the magnetization of the system. The relative frequency of the magnetization jumps $\Delta M$ is reported in Fig.~\ref{fig:fig03} by using log-log axes, and a proper logarithmic binning. The $x$-axis displays the amplitude $\Delta M$ of the avalanche, while the $y$-axis reports the probability of jumps with size $\Delta M$ falling into each bin, whose size is chosen according to a standard logarithmic binning procedure. The presence of a power-law distribution appears as a linear plot over several decades. As shown in Fig.~\ref{fig:fig03}, close to the transition $T_c(R)$, a power-law distribution emerges which spans several decades. At any given $T$, the maximum extension is obtained at the critical disorder determined as discussed above. However, a power-law distribution is observed over several decades also slightly off the transition. This is a key result, as the disorder $R$ is an uncontrolled feature in a real system, which is not necessarily expected to stay at its critical value. Clearly, a complete power-law distribution is expected to emerge only at the critical value of the disorder, but we get avalanches also off the critical point. As expected, supra-critical values of $R$ (black points in Fig.~\ref{fig:fig03}) evidence a large-avalanche cutoff, while sub-critical disorder values (see, \emph{e.g.} the orange points) record an over presence of large avalanches. Finally, it is interesting to notice that the (portions of) power-law distributions recorded at any $R$ share quite similar values for the slope, which suggests a weak dependence of the critical power-law exponent from the disorder parameter.
	
\begin{figure}
		\begin{center}
			\includegraphics[width=0.49\textwidth]{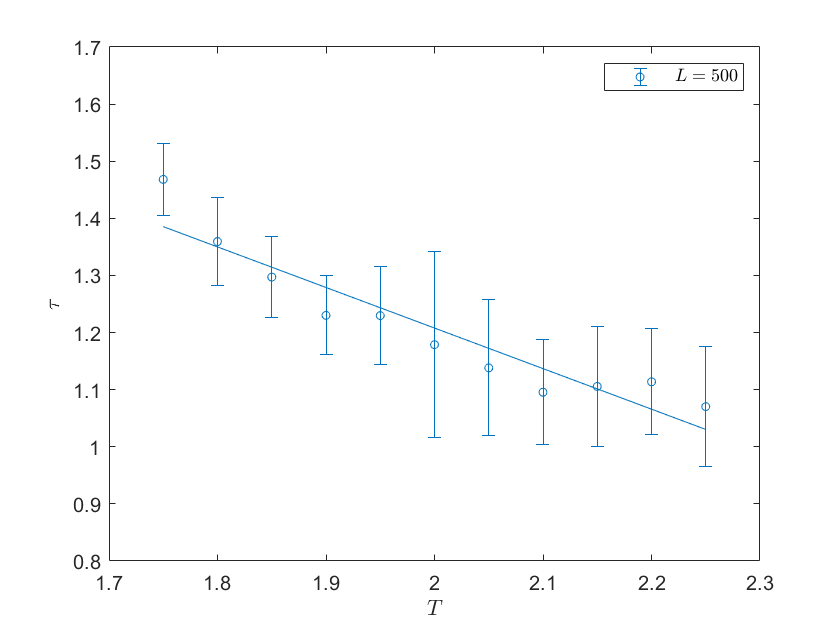}
			\includegraphics[width=0.49\textwidth]{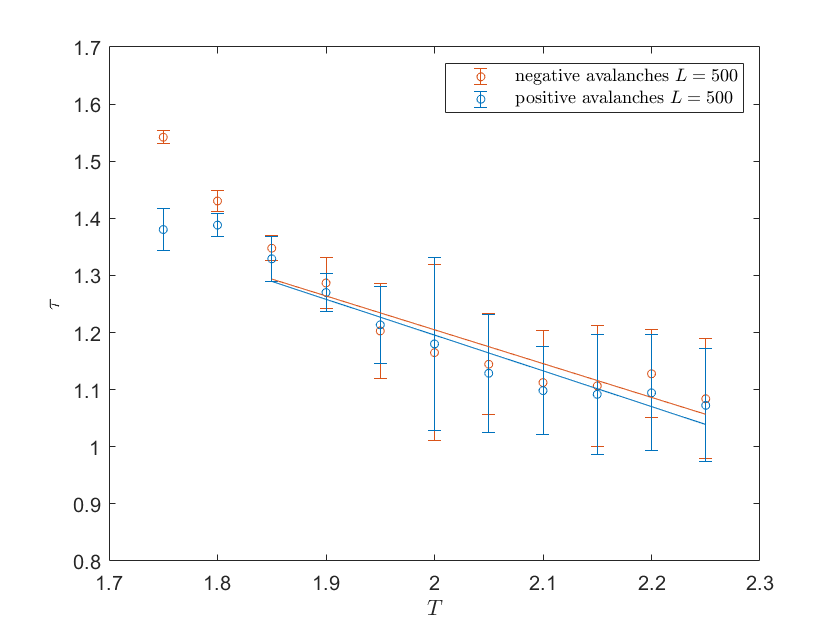}
		\end{center}
		\caption{Left: Critical exponent of the power-law distribution derived for the Barkhausen jumps along the critical line $T_\text{c}(R)$. All the reported data are obtained for a $L\times L=500\times 500$ spin system. Error bars derived considering subcritical and critical distributions. Right: Results for the critical exponents obtained by separating positive and negative magnetization jumps in critical distributions.}
		\label{fig:TauVsT}
	\end{figure}

	For each temperature the avalanche statistics identify a critical exponent which is obtained by a linear fit of the log-log plots. This critical exponent varies as we move along the $T_\text{c}(R)$ critical line, and its dependence is shown in the left panel of Fig.~\ref{fig:TauVsT}. This behaviour is in quite remarkable agreement with the experimental data obtained in \cite{00Puppin,04Puppin}. We also tested whether the negative and the positive magnetization avalanches share the same statistical properties, a fact that was experimentally observed in \cite{03Zani}. The right panel of Fig.~\ref{fig:TauVsT} confirms that this is the case in a wide range of temperatures. All these results point in the direction that the underlying microscopic mechanism responsible for the Barkhausen effect might be the presence of a certain degree of randomness in the interaction strength among the neighbouring spins.
	
	\section{Defects-induced magnetization jumps}\label{sec:def}
	
	In \cite{22esb} the reversal transition of a magnetic system with pinned defects has been analysed. The presence of defects triggers a number of effects, including a reduction of the critical area for nucleating droplets and of the metastable lifetime, as well as a shift of the dynamic spinodal line which ensures that smaller systems are able to undergo homogeneous multi-droplet transformations. We here pursue further the study, and analyse the sequence of magnetization jumps that emerge during a magnetization reversal induced by an external field.
	
	We now consider a 2D Ising system composed of $L\times L$ spins, evolving under the effect of the Hamiltonian \eqref{eq:Hamiltonian} with periodic conditions applied at the boundaries. In order to model the presence of pinned defects, a fraction $f$ of the spins is held fixed. The defects are quenched in random positions, with a fraction $f/2$ of them pointing in each spin direction. Thus, defects aligned with an external field induce nucleation of a coherent domain, while those opposed to the field hamper the domain growth. The system is initialised with all spins parallel to the external field. After an initial thermalisation time, the field starts oscillating with a period $P$ which ensures that the spins are able to follow the magnetization reversal, so that the system is in the disordered side of the dynamic phase transition \cite{08buendia,20mar}. The evolution is modeled through the rejection-free Kinetic Monte-Carlo algorithm described in \cite{22esb}. This implements the Glauber dynamics, and therefore we are able to associate a (move-dependent) time interval to each individual single-flip move. More precisely, the $i$-th spin flip has the transition probability rate
	\begin{equation}\label{eq:wi}
		w_i=\frac{1}{2\alpha}\,(1-s_i\tanh\beta h_i),
	\end{equation}
	where $\alpha$ is a microscopic time (henceforth set equal to unity), and $h_i$ is the local field on the $i$-th spin. If we set $w_\text{T}=\sum_i w_i$, the time interval associated with each chosen move is extracted from a Poissonian with parameter $w_\text{T}^{-1}$. As a result, the more the system is out-of-equilibrium (and moves exist that significantly reduce the energy), the shortest time interval is associated with the next move.
	
	\subsection{Measurement time and power-law distributions}
	
	In order to define and measure the magnetization jumps we must specify a time interval $\Delta t_\text{meas}$, the \emph{measurement time}. This defines a set of specific times $t_n=t_0+n \Delta t_\text{meas}$, and the magnetization jumps $\Delta M$ will now be defined as the magnetization variation between each $t_n$ and the previous $t_{n-1}$. We remark that since our Monte-Carlo algorithm implements the Glauber dynamics, the number of Monte-Carlo steps included in a chosen $\Delta t_\text{meas}$ is not constant, especially when the simulated experiment visits both \emph{quasi}- and out-of-equilibrium configurations. Magnetization avalanches are then identified by gathering in a single, larger magnetization jump all subsequent single jumps which share the same sign of $\Delta M$.
	
	The choice of $\Delta t_\text{meas}$ is crucial. A measurement time too small generates magnetization variations dominated by thermal fluctuations, so that a very large number of small jumps of each sign is recorded. On the contrary, a too large $\Delta t_\text{meas}$ becomes blind to any sort of fluctuations, and we end up measuring a limited number of (too large) jumps.
	
\begin{figure}
		\begin{center}
			\includegraphics[width=0.75\textwidth]{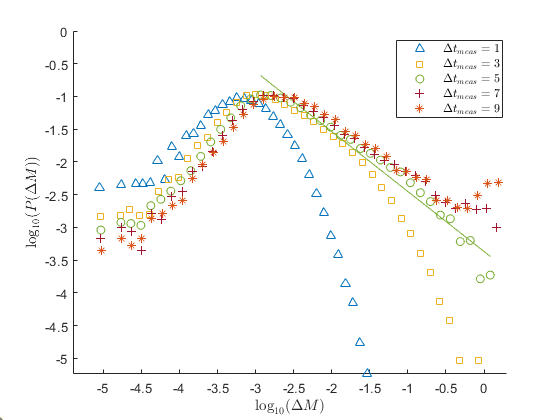}
		\end{center}
		\caption{Magnetization jumps distributions obtained by varying the measurement time $\Delta t_\text{meas}$. The plots are obtained by simulating a system composed by $500\times 500$ spins, with a defect fraction of 2\%, monitored at temperature $T=1$, under the effect of an oscillating magnetic field with peak intensity $B=1$ and period $P=10^6$. The distributions correspond to the measurement times $\Delta t_\text{meas}$ displayed in the inset. A line with unit slope is displayed close to the green points (corresponding to $\Delta t_\text{meas}=5$) for the reader's ease.}
		\label{fig:taum}
	\end{figure}

	Fig.~\ref{fig:taum} confirms the above by reporting the results of a series of simulations performed on a $500\times500$ spins system, with different choices of the measurement time $\Delta t_\text{meas}$. In order to avoid recording the fluctuations about an equilibrium configuration, we monitored the magnetization jumps only in the transition regime, defined as the set of configurations in which $-0.9<M<+0.9$, since the equilibrium magnetization under the effect of the peak magnetic field is larger than 0.95 (in modulus). All the reported data exhibit an initial, increasing portion of the jumps distribution, corresponding to small jumps, followed by a heavy tailed regime. The fact that all the distributions exhibit a low-jumps cutoff is due both to the finite size of the system and to the choice of the measurement time: the longer $\Delta t_\text{meas}$, the less likely small jumps are. When the measurement time is chosen optimally (intermediate set of data, displayed as green circles), the heavy tail is a power law which expands over three decades.
	
	\subsection{Temperature and randomness} \label{sec:TadnR}
	
	Varying the temperature and/or the defect fraction has a qualitatively similar effect on the bursty magnetization jumps. A large temperature or a large fraction of quenched defects makes it particularly difficult to generate and/or reverse large domains. Small domains or even individual spin flipping become more and more present, with a net effect that a large-jumps cutoff is to be expected. On the other hand, if the temperature is too low and/or the system becomes homogeneous, macroscopic domains tend to form and the distributions exhibit a predominance of very large jumps.
	
	\begin{figure}
		\begin{center}
			\includegraphics[width=0.49\textwidth]{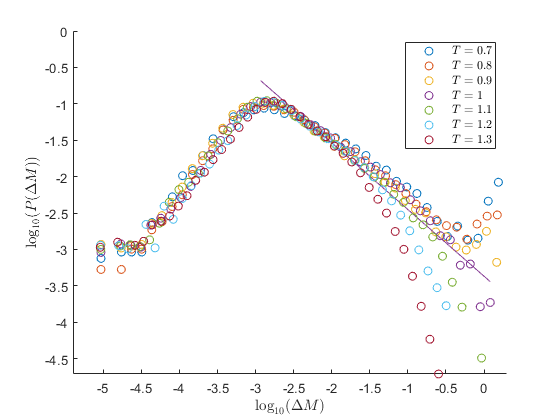}
			\includegraphics[width=0.49\textwidth]{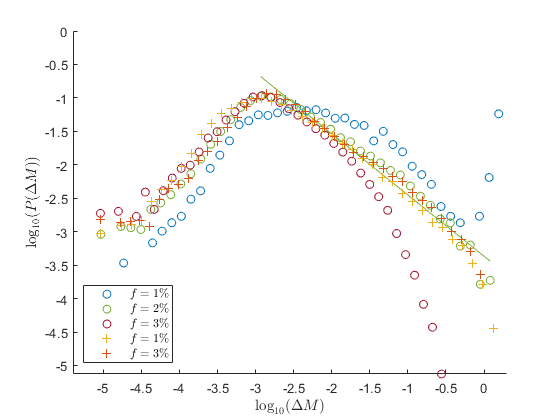}
		\end{center}
		\caption{Magnetization jumps distributions obtained by varying the temperature (left) or the defect fraction (right). All data are collected for a system as in Fig.~\ref{fig:taum}, monitored with a $\Delta t_\text{meas}=5$. Left: distributions corresponding to different temperatures, for $f=2\%$. Right: distributions corresponding to different defect fractions, for $T=1.0$. The crosses report also the distributions obtained when we choose the optimal temperatures and measurement times for different defect fractions. Precisely: green circles correspond to $f=2\%$, with $T=1$ and $\Delta t_\text{meas}=5$; yellow crosses correspond to $f=1\%$, with $T=0.9$ and $\Delta t_\text{meas}=1$; red crosses correspond to $f=3\%$, with $T=1.0$ and $\Delta t_\text{meas}=10$.}
		\label{fig:tandf}
	\end{figure}
	
	Fig.~\ref{fig:tandf} confirms and quantifies the above expectations. From the left panel, it is noteworthy to remark that the temperature influences almost exclusively the right tail of the distributions, thus impacting the number and magnitude of larger jumps, while the small activity proceeds almost insensitive to temperature variations. In the right panel, and besides the data corresponding to different defect fractions, monitored at the same temperature, we have also reported the distributions that can be obtained by varying the defect fraction but, for each value of $f$, choosing the optimal values of temperature and measurement time in terms of obtaining the longest power-law. It is striking that the three data (corresponding to $f=1\%$, 2\%, or 3\%) share a quite similar power-law distribution at their critical temperatures, and it is interesting to notice that, within the error bars, the power laws fit well to the reported straight line corresponding to a critical exponent $\tau=1$.
	
	\section{Discussion}
	\label{sec:conclusion}
	
	\subsection{Finite size analysis}
	
	Before discussing the results we first report the analysis of the finite-size effects on the jumps distributions we measure, to check whether the measured power-laws exponents provide reliable estimates of the expected distributions at the thermodynamic limit. We have thus simulated smaller systems (with $L=50$, 100, or 200) and derived the associated avalanche statistics. The results thus obtained are here compared with the previous results, obtained with $L=500$.
	
\begin{figure}
		\begin{center}
			\includegraphics[width=0.49\textwidth]{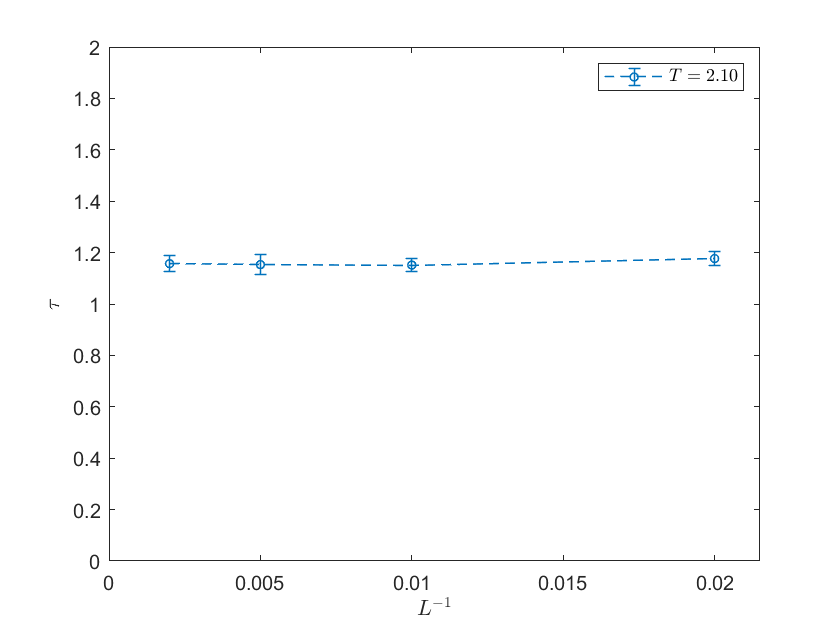}
			\includegraphics[width=0.49\textwidth]{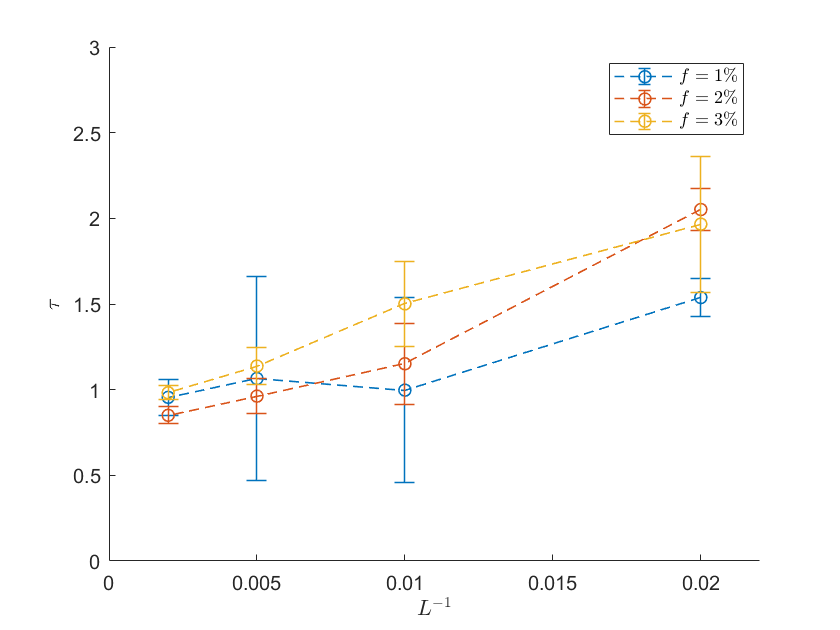}
		\end{center}
		\caption{Critical exponent function of the inverse of the system size $L$. Left: the cluster method case, for fixed temperature $T=2.10$ and a subcritical randomness $R=0.23$. Error bars are obtained by performing 3 independent analysis on systems with slightly different $R=0.22,0.23,0.24$. Right: magnetic defects case, for different fraction of defects $f=1\%,2\%,3\%$. Error bars reproduce the standard deviation considering the power law exponents extracted from 10 independent realization of the system randomness.}
		\label{fig:FiniteSize}
	\end{figure}

	For the random-bond Ising model (RBIM), the left panel of Fig.~\ref{fig:FiniteSize} reports the critical exponents obtained for systems of different sizes, simulated at a fixed temperature $T=2.10$ and the associated critical value of the randomness parameter. Error bars in the panel are obtained by considering three values of $R$ in a small interval centred in $R_c$. It comes to no surprise to report that smaller systems exhibit power laws spanning over less orders of magnitude. Nevertheless, Fig.~\ref{fig:FiniteSize} evidences that the critical exponent $\tau$ does not show a remarkable system size dependence, so we can be confident with the results obtained for a system with size $L=500$.
	
	In the presence of magnetic defects, we performed simulations with fractions of defects $f=1\%$, 2\%, and 3\%. For each of these values, we selected the temperature and measurement time fit to retain the longest power law distribution, and we studied systems of variable size ($L=50$, 100, or 200, besides the case $L=500$, already reported). The results obtained for the associated critical exponents are reported in the right panel of Fig.~\ref{fig:FiniteSize}. The error bars are derived by considering the value of $\tau$ in 10 independent realizations of the randomness in the system. Fig.~\ref{fig:FiniteSize} evidences that smaller systems exhibit a larger oscillation of the critical exponent, especially if the defect fraction increases. However, when $L=500$ the system-to-system variation (and therefore the error bars) is strongly reduced, and all data are consistent with the estimate $\tau\approx 1$.
	
	\subsection{Conclusions}
	
	We have studied the Barkhausen noise emerging in two different random variants of the Ising model. We first focused in the random-bond Ising model (RBIM) and identified the parameter values for which the system is in the ferromagnetic state. The disorder parameter is defined as the variance of the probability distribution from which the random interaction strengths are derived. At any finite temperature, small values of the disorder parameter induce a coherent response in which large jumps dominate over medium and intermediate domains reversal. On the contrary, when the disorder parameter attains too large values, small domains evolve independently and prevent the formation of larger avalanches. In between, a critical regime can be identified for any temperature (see Fig.~\ref{fig:fig03}). We also studied the temperature-dependence of the critical exponents characterising (see Fig.~\ref{fig:TauVsT}) the emerging power-law distributions. The critical exponent increases when temperature decreases, and positive and negative avalanches both share the same critical exponents, all in agreement with experimental observations \cite{00Puppin,03Zani,04Puppin}.
	
	Randomness might also be induced in an Ising system by quenching in random positions a finite portion of defects (\emph{i.e.}, spin which never reverse). The magnetic response of a system thus prepared can be then be examined under the effect of a slowly varying external field. By simulating the system behaviour with the Glauber dynamics we have been able to enlighten the key role played by the measurement time
	(see Fig.~\ref{fig:taum}). More precisely, if we record the magnetization values too often, thermal fluctuations might hinder the bursty dynamics characteristic of the Barkhausen effect. On the contrary, if we wait too long until recording the magnetization, the system finds its way to complete the reversal transition. In between a critical range of measurement times can be identified, for which power-law distributions characterise the Barkhausen jumps. Fig.~\ref{fig:tandf} quantifies the effect of the defect fraction, which is quite similar to the disorder parameter in the RBIM. However, at variance with the RBIM observations, the exponents we found when the critical defect fraction is chosen are all quite close to the unique value $\tau=1$, which suggests that both random variants of the Ising model belong to different universality classes.
	
	\section*{Declarations}
	The authors have no relevant financial or non-financial interests to disclose.
	\noindent The datasets generated and analysed during the current study are available from the corresponding author on reasonable request.
	
	\bibliography{sn-bibliography}
	\bibliographystyle{elsarticle-num}
	
\end{document}